\documentclass[twocolumn,pra,amsmath,showpacs]{revtex4}
\usepackage{graphicx}
\begin{document}
\title{Classifying $N$-qubit Entanglement via Bell's Inequalities}
\author{Sixia Yu$^1$, Zeng-Bing Chen$^1$, Jian-Wei Pan$^{1,2}$, and Yong-De Zhang$^1$}
\affiliation{$^1$Department of Modern Physics, University of
Science and Technology of China, Hefei 230027, P.R. China}
\author{ }
\affiliation{$^2$Institut f\"ur Experimentalphysik, Universit\"at
Wien, Boltzmanngasse 5, 1090 Wien, Austria}
\date{\today}

\begin{abstract}
All the states of $N$ qubits can be classified into $N-1$
entanglement classes from 2-entangled to $N$-entangled (fully
entangled) states. Each class of entangled states is characterized
by an entanglement index that depends on the partition of $N$. The
larger the entanglement index of an state, the more entangled or
the less separable is the state in the sense that a larger maximal
violation of Bell's inequality is attainable for this class of
state.
\end{abstract}
\maketitle

Bell's inequalities \cite{bell} were initially  dealing with two
qubits, i.e., two-level systems. They ruled out various kinds of
local hidden variable theories. Recently with the emergence of the
field of quantum information where the entangled states are
essential, Bell's inequalities provide also a necessary criterion
for the separability of 2-qubit states. This is because Bell's
inequalities are observed by all  separable 2-qubit states. For
pure states Bell's inequalities are also sufficient for
separability \cite{gisin1}. The more entangled the state, the
larger is the maximal violation of Bell's inequalities.

On the one hand, Bell's inequalities were generalized to $N$
qubits \cite{mermin, gisin,bruk}, whose violations provide a
criterion to distinguish the totally separable states from the
entangled states. On the other hand with the experimental
realization of multiparticle entanglement \cite{pan}, Bell's
inequalities in terms of the Mermin-Klyshko (MK) polynomials
\cite{mermin,kly} were generalized to the case of $N=3$  for the
detection of fully entangled 3-qubit states \cite{svet1} and to
the $N$-qubit case for the detection of  fully entangled $N$-qubit
states\cite{collins,svet2}. And these results about full
entanglement are inferred by the quadratic Bell inequalities
\cite{uffink}.

In this Letter we shall provide a detailed classification of
various types of $N$-qubit entanglement from total separability to
full entanglement. It turns out that an entanglement index can be
defined to characterize the entanglement class. Before presenting
our classification of $N$-qubit entanglement we shall at first
review the classification of 2-qubit and 3-qubit entanglement by
Bell's inequalities.

{\it Quadratic Bell's inequalities for 2-qubit system---}At first
let us consider a system of two qubits labelled by $A$ and $B$.
There are only two types of states, separable or entangled. If a
2-qubit state $\rho$ is separable, i.e., a pure product state or a
mixture of pure product states, the well-known Bell-CHSH (Clauser,
Horne, Shimony, and Holt) inequality \cite{CHSH}
 \begin{equation}\label{bi1}
 \langle AB+AB^\prime+A^\prime B-A^\prime B^\prime\rangle_\rho\le 2
 \end{equation}
holds true for all testing observables $A^{(\prime)}=\vec
a^{(\prime)}\cdot\vec \sigma_A$ and $B^{(\prime)}=\vec
b^{(\prime)}\cdot\vec\sigma_B$. Here $\sigma_A$ and $\sigma_B$ are
the Pauli matrices for qubits $A$ and $B$ respectively; the norms
of the real vectors $\vec a^{(\prime)}$, $\vec b^{(\prime)}$ are
less than or equal to 1; $\langle AB\rangle_\rho={\rm Tr}(\rho
AB)$ denotes the average of the observable $AB$ in the state
$\rho$ as usual.

If the state $\rho$ is entangled, then the upper bound of the
average of the Bell operator is $2\sqrt2$ \cite{sqrt2}, which is
attainable for maximally entangled states. For the present purpose
this upper bound is expressed more properly in a quadratic form of
Bell's inequality \cite{uffink}
 \begin{equation}\label{uf}
 \langle X\rangle_\rho^2+ \langle Y\rangle_\rho^2\le 4
 \end{equation}
in terms of two observables $X=A^\prime B+AB^\prime$ and
$Y=AB-A^\prime B^\prime$ instead of one Bell operator $X+Y$. This
inequality is satisfied by all 2-qubit states, entangled or
separable. For separable states the Bell-CHSH inequality
(\ref{bi1}) can also be equivalently expressed  as
 \begin{equation}\label{bi3}
 |\langle X\rangle_\rho|+ |\langle Y\rangle_\rho|\le 2
 \end{equation}
in terms of observables $X$ and $Y$  because of the arbitrariness
of the testing observables $A^{(\prime)}$ and $B^{(\prime)}$.
\begin{center}
\begin{figure}[h]
\includegraphics{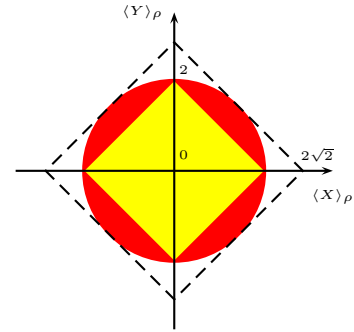}
\caption{All separable states lie in the inner square
 $|\langle X\rangle_\rho|+ |\langle Y\rangle_\rho|\le 2$ while a general 2-qubit state, whether separable or not,
 is bounded by a circle $\langle X\rangle_\rho^2+\langle Y\rangle_\rho^2\le 4$ instead of the dashed square
  $|\langle X\rangle_\rho|+ |\langle Y\rangle_\rho|\le 2\sqrt2$.}
 \end{figure}\vskip -1cm
\end{center}

To summarize, there is only one entanglement class of 2-qubit
states, i.e., 2-entangled states. For the separable states the
Bell-CHSH inequality is observed while for the entangled states
the Bell-CHSH inequality can be violated. By regarding $\langle
X\rangle_\rho$ and $\langle Y\rangle_\rho$ as two axes of a plane
respectively, we can put all the results known so far in a diagram
as shown in Fig.1.

{\it Classification of 3-qubit entanglement---}Now we consider
three qubits labelled by $A,B$ and $C$. There are three types of
3-qubit states: i) totally separable states denoted as
$(1_3)=\{$mixtures of states of form
$\rho_A\otimes\rho_B\otimes\rho_C\}$; ii) 2-entangled states which
is denoted as $(2,1)=\{$mixtures of states of form $
\rho_A\otimes\rho_{BC}, \rho_{AC}\otimes\rho_B,
\rho_C\otimes\rho_{AB}\}$; iii) fully entangled states which is
denoted as $(3)=\{\rho_{ABC}\}$ including the
Greenberger-Horn-Zeilinger (GHZ) state \cite{ghz}.

These three types of 3-qubit states can be discriminated from one
another by the averages of the MK polynomials defined as:
 \begin{subequations}\begin{eqnarray}
 F_3&=&(AB^\prime +A^\prime B)C+(AB-A^\prime B^\prime) C^\prime,\\
 F_3^\prime&=&(AB^\prime +A^\prime B)C^\prime-(AB-A^\prime B^\prime)
 C,
 \end{eqnarray}\end{subequations}
where $C$ and $C^{\prime}$ are two observables of the third qubit
defined similarly to $A$ and $B$. In fact for totally separable
states the Bell-Klyshko inequality reads
 \begin{equation}\label{c33}
 \max\{|\langle F_3\rangle_\rho|, |\langle F_3^\prime\rangle_\rho|\}\le 2
 \quad \mbox{if} \quad \rho\in (1_3),
 \end{equation}
whose violation ensures a nonseparable state which can be either a
2-entangled state or a fully entangled state. The maximal
violation is different for 2-entangled states and fully entangled
states \cite{uffink}
 \begin{subequations}
 \begin{eqnarray}
 \langle F_3\rangle_\rho^2+\langle F_3^\prime\rangle_\rho^2&\le 2^3 \quad \mbox{if}& \rho\in (2,1)\label{c31},\\
 \langle F_3\rangle_\rho^2+\langle F_3^\prime\rangle_\rho^2&\le 2^4 \quad \mbox{if}& \rho\in (3)\label{c32}.
 \end{eqnarray}
 \end{subequations}
Therefore the violation of inequality Eq.(\ref{c31}) ensures a
fully entangled state.
\begin{center}\vskip -0.2cm
\begin{figure}[h]
\includegraphics{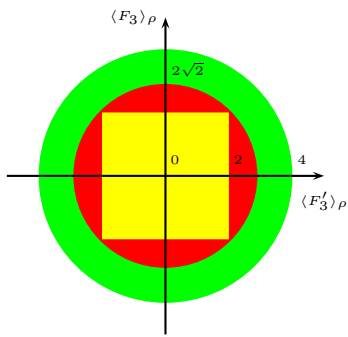}
\caption{Classification of 3-qubit entanglement. Separable states
 reside inside the square while the 2-entangled states are bounded
 by the smaller circle. If the averages of $F_3$ and $F_3^\prime$ in one state
  are outside the smaller circle then the state is fully entangled.}
 \end{figure}\vskip-1.1cm
\end{center}%

In summary, there are two different entanglement classes of
3-qubit states, namely, 2-entangled states and fully entangled
states. Every class of entangled states can give rise to different
violation of the Bell-Klyshko inequality. We can put all these
known results into a diagram as Fig.2 by regarding the averages of
observables $F_3$ and $F_3^\prime$ as two axes of a plane. Since
the shape of the diagram is like a piece of ancient Chinese coin
so this kind of diagram will be referred to as an ACC diagram.

{\it Classification of $N$-qubit entanglement---}Now we turn to
the classification of $N$-qubit states. The number of the types of
$N$-qubit states is the same with the number of all partitions
$(\vec n)= (n_1,n_2,\ldots,n_N)$ of $N$ with $n_i$
$(i=1,\ldots,N)$ being integers such that
 \begin{equation}
 \sum_{i=1}^Nn_i=N, \quad (N\ge n_1\ge n_2\ge\ldots\ge n_N\ge0).
 \end{equation}
In fact the partition $(\vec n)$ is in a one-to-one correspondence
with the types of states that are mixtures of states
$\rho_{n_1}\otimes\rho_{n_2}\otimes\cdots\otimes\rho_{n_N}$, where
$\rho_{n_i}$ is a fully entangled state of any $n_i$ qubits
$(i=1,\ldots,N)$. Therefore we can label different types of
$N$-qubit states by different partitions of $N$. For example, the
partition $(N)$ corresponds to the fully entangled states while
the partition $(1,1,\ldots,1)=(1_N)$ corresponds to the totally
separable states. As a result the number of different types of
$N$-qubit states is also the number of all the irreducible
representations of the permutation group with $N$ elements. This
is not a coincidence because the types of the states as well as
the MK polynomials defined below are invariant under the
permutations of qubits.

To classify all types of $N$-qubit states, we shall employ the MK
polynomials for the system of $N$-qubit. If we define $F_2=X+Y$
and $F_2^\prime=X-Y$ the MK polynomials are defined recursively as
 \begin{equation}\label{fn1}
 F_N=\frac12(D_N+D_N^\prime)F_{N-1}+\frac12(D_N-D_N^\prime)F_{N-1}^\prime
 \end{equation}
for $N\ge3$ where $D_N$ and $D_N^\prime$ are observables of the
$N$-th qubit and $F_N$ and $F_N^\prime$ are MK polynomials for
$(N-1)$-qubit. And observable $F_N^\prime$ is defined similarly to
$F_N$ with the primed and unprimed observables interchanged. The
averages of these two observables in totally separable states
satisfy the Bell-Klyshko inequality
 \begin{equation}
 \max\{|\langle F_N\rangle_\rho|,|\langle F_N^\prime\rangle_\rho|\}\le 2\quad\mbox{if}\quad \rho\in(1_N).
 \end{equation}

As we will see immediately some types of states are more entangled
than other types in the sense that a larger violation of this
inequality is attainable. And the upper bound of the violation of
the state in $(\vec n)$ is related to {\em the entanglement index}
defined as
 \begin{equation}
 E_N(\vec n)=N-K_1(\vec n)-2L(\vec n)+2,
 \end{equation}
where $L(\vec n)$ is the number of entries in $(\vec n)$ that are
greater than or equal to 2 and $K_1(\vec n)$ is the number of
entries in $(\vec n)$ that equal to 1, i.e., the number of
separated single qubits. In other words, $N-K_1(\vec n)$ is
exactly the number of entangled qubits while $L(\vec n)$ is
exactly the number of groups into which the entangled $N-K_1$
qubits are divided with each group of qubits fully entangled.
Obviously the entanglement index is an integer satisfying $2\le
E_N(\vec n)\le N$.

For example, in the case of $N=4$ we have 5 partitions, therefore
5 types of states: i) fully entangled 4-qubit states $(4)$ with
$L(4)=1$, $K_1(4)=0$ and $E_4(4)=4$; ii) states of a group of
fully entangled 3-qubit and a separated single qubit $(3,1)$ with
$L(3,1)=1$, $K_1(3,1)=1$ and $E_4(3,1)=3$; iii) $(2,2)$ stands for
the states of two groups of  entangled 2-qubit with $L(2,2)=2$,
$K_1(2,2)=0$ and $E_4(2,2)=2$; iv) $(2,1,1)=(2,1_2)$ corresponds
to the state of an entangled 2-qubit together with two separated
qubits with $L(2,1_2)=1$, $K_1(2,1_2)=2$ and $E_4(2,1_2)=2$; v)
$(1,1,1,1)=(1_4)$ corresponds to the totaly separable states with
$L(1_4)=0$, $K_1(1_4)=4$, and $E_4(1_4)=2$.

According to the entanglement index all the $N$-qubit states can
be classified into $N$ entanglement classes: the class of totally
separable states $S_1$ and the class of {\em $E$-entangled states}
$\rho_E\in S_E$ $(E=2,\ldots,N)$, which are mixtures of entangled
states with the same entanglement index $E$
 \begin{equation}
 \rho_E=\sum_{E_N(\vec n)=E}p({\vec
 n})\;\rho_{n_1}\otimes\rho_{n_2}\otimes\cdots\otimes\rho_{n_N},
 \end{equation}
where the summation is over all states corresponding to the same
partition and all partitions with the same entanglement index $E$
and $p(\vec n)$ is a probability distribution.  One can also
define a separability index as $S_N(\vec n)=2L(\vec n) +K_1(\vec
n)$ with $S_N+E_N=N+2$. An $E$-entangled state is also an
$S$-separable state. The separability index $S_N$ can be regarded
as the effective number of qubits that can realize the partition
$(\vec n)$ since there are at least two qubits in each groups of
fully entangled qubits. Now we are ready to present our main
result.

{\em Classification Theorem: The $N$-qubit states are classified
into $N-1$ entanglement classes and the $E$-entangled states
satisfy the following quadratic Bell inequality
 \begin{equation}
 \langle F_N\rangle_\rho^2+\langle F_N^\prime\rangle_\rho^2
 \le 2^{E+1}\quad \mbox{if}\quad\rho\in S_E.
 \end{equation}
Therefore the larger the entanglement index, the larger is the
maximal violation of the Bell-Klyshko inequality.  In this sense
the larger the entanglement index, the more entangled or the less
separable is the state.}

We shall postpone the proof to the next section. The upper bound
of the inequality for the $E$-entangled states is attainable for
the pure state of type $(n_1,n_2,\ldots,n_N)$ that is a product of
maximal entangled states for $n_i$ qubits where for $n_i\ge3$ the
maximal entangled state is chosen as the GHZ state and for $n_i=2$
the maximal entangled state is chosen as one of the Bell states.

The largest entanglement index $E_N(N)=N$ is reached by the
partition $(N)$. Therefore the $N-$entangled states or fully
entangled states can be distinguished from other classes of states
by Bell's inequality. The $(N-1)$-entangled states are states of
one group of entangled $N-1$ qubits and a separated single qubit.
The $(N-2)-$entangled states are states of a group of fully
entangled $(N-k)$-qubit and a group of fully entangled $k$-qubit
with $k\ge 2$. Thus the states with different $k$ have the same
property of non-separability. The least entanglement index is
possessed by the 2-entangled states.

We notice that when $K_1=0$ there is no separated qubits and the
more entries in the partition $(\vec n)$, i.e., the more groups
into which the entangled qubits are divided, the more separable is
the state. Conversely for states with larger entanglement index,
the $N$ qubits are divided into less groups of fully entangled
qubits.

One may tend to think that the more qubits are involved in the
entanglement the larger the violation of Bell's inequality and
therefore less separable is the state. However this is not always
true. In the case of 10 qubits, the states of type (5,2,2,1)  will
have a smaller entanglement index than the states of type (4,3,3).
Namely we have $E_{10}(5,2,2,1)=6$ and $E_{10}(4,3,3)={7}$. Thus
the state of type (5,2,2,1) is a 6-entangled state and the state
of type (4,3,3) is a 7-entangled state. In other words, the former
is more separable than the latter in spite of the 5-qubit
entanglement of the former. Therefore the entanglement index
indicates a global property of how the $N$-qubit entanglement is
distributed among $N$ qubits rather than local property of how
entangled are the separated groups of fully entangled qubits.

The classification of $N$-qubit entanglement can also be put into
an ACC diagram as in Fig.3 by regarding the average of $F_N$ and
$F_N^\prime$ as two axes of a plane. Here $N-1$ circles and one
square are the boundary for all kinds of $N$-qubit entanglement.
\begin{center}
\vskip -.5cm
\begin{figure}[h]
\includegraphics{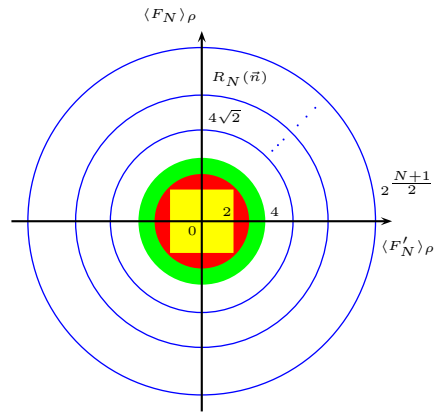}
 \caption{Classification of $N$-qubit entanglement. While all separable states
 are bounded in the square, the states of type $(\vec n)$ are
 bounded within a circle of a radius $R_N=2^{(E_N(\vec n)+1)/{2}}$.
 The largest circle with a radius of
 $2^{(N+1)/2}$ corresponds to the fully entangled states.}
 \end{figure}\vskip -1.2cm
\end{center}

{\it Proof of the theorem---}At first we notice that the theorem
is true when $N=3$ where we have $E_3(3)=3$ and  $E_3(2,1)=2$.
Since an $E$-entangled state $\rho_E$ of $N$ qubits is a convex
mixture of states in $(\vec n)$ with the entanglement index
$E_N(\vec n)=E$, it is enough to prove the inequality in the
theorem for a general state $\rho$  corresponding to the partition
$(\vec n)$. For convenience we shall denote ${\mathcal B}_N(\vec
n)=\langle F_N\rangle_\rho+\langle F_N^\prime\rangle_\rho$.
Further if the smallest entry of partition $(\vec n)$ is $k\ge1$
then we denote $(\vec n)=(\vec n_k,k)$, where $(\vec n_k)$ is a
partition of $N-k$ whose smallest elements is greater or equal to
$k$.

If the state $\rho$ is of type $(\vec n_1,1)$ then there is at
least one separated single qubit. Because the MK polynomials are
symmetric under the permutation of qubits, it is enough to suppose
that the $N$-th qubit is separated. From the definition
Eq.(\ref{fn1}) of the MK polynomials we obtain
 \begin{equation}\label{b1}
 {\mathcal B}_N(\vec n_1,1)=\frac12(\langle D_N\rangle_\rho^2+\langle D_N^\prime\rangle_\rho^2){\mathcal B}_{N-1}(\vec
 n_1)\le{\mathcal B}_{N-1}(\vec n_1).
 \end{equation}
Obviously this inequality is attainable. If the state $\rho$ is of
type $(\vec n_k,k)$ with $k\ge 2$, then there is at least one
group of fully entangled k-qubit. Without loss of generality we
suppose the last $k$ qubits are separated from other $N-k$ qubits.
By rewriting the MK polynomial as \cite{gisin}
 \begin{equation}
 F_N=\frac14(F_{N-k}(F_k+F_{k}^\prime)+F_{N-k}^\prime(F_k-F_{k}^\prime)),
 \end{equation}
where $F_k$ and $F_k^{\prime}$ are MK polynomials for the last $k$
qubits and $F_{N-k}$ and $F_{N-k}^{\prime}$ are MK polynomials for
the rest $N-k$ qubits we obtain
 \begin{equation} \label{b3}
 {\mathcal B}_N(\vec n_k,k)=\frac18{\mathcal B}_k(k){\mathcal B}_{N-k}(\vec n_k)\le
2^{k-2}{\mathcal B}_{N-k}(\vec n_k).
 \end{equation}
where we have used the quadratic Bell inequality for the fully
entangled $k$-qubit state ${\mathcal B}_k(k)\le 2^{k+1}$
\cite{uffink}, which is attainable for the GHZ state for $k$
qubits. Therefore the above inequality is also attainable.

By denoting $K_k$ as the number of $k's$ as entries of the
partition $(\vec n)$, we can rewrite the partition as $(\vec
n)=(M_{K_M},\ldots,2_{K_2},1_{K_1})$ where $M$ is the largest
entry of $(\vec n)$. As a result \begin{widetext}
 \begin{eqnarray}
 {\mathcal B}_N(\vec
 n)&=&{\mathcal B}_N(M,M_{K_M-1},(M-1)_{K_{M-1}},\ldots,3_{K_3},2_{K_2},1_{K_1})\cr
 &\le&2^{K_3(3-2)}2^{K_4(4-2)}\cdots 2^{K_{M-1}((M-1)-2)}2^{(K_M-1)(M-2)}{\mathcal B}_M(M)\cr
 &\le&2^{3+\sum_{k=3}^MK_k(k-2)}=2^{N+3-2L-K_1}=2^{E_N(\vec n)+1},
 \end{eqnarray}
\end{widetext}
where identities $N=\sum_{k=1}^MkK_k$ and $L=\sum_{k=2}^MK_k$ have
been used. The first inequality is a consequence of applying
inequality Eq.(\ref{b1}) $K_1$ times and Eqs.(\ref{b3}) $K_k$
times for $k=3,4,\ldots,M-1$ and $K_M-1$ times when $k=M$. The
second inequality stems from the quadratic Bell's inequality for
fully entangled $M$-qubit states. Since all the inequalities in
the proof are attainable, the upper bound for the $E$-entangled
states is also attainable.

{\it Conclusions and discussions---}In this Letter we have
provided a classification of $N$-qubit entanglement by introducing
the integer entanglement index $E_N$. Some orders have been
established among various types of $N$-qubit entanglement: the
larger the entanglement index, the larger maximal violation of the
Bell's inequality is attainable and therefore more entangled or
less separable is the state.

However our classification is far from being complete regarding to
the following two connected aspects. First, as Bell's inequality
provides only a sufficient criterion for the entanglement, the
hierarchy of quadratic Bell inequalities for different classes of
entangled states is sufficient for the detection of
$E$-entanglement. Second, within one entanglement class there may
exist states with different types of entanglement. These
differences obviously cannot be detected by Bell's inequalities in
terms of MK polynomials discussed here. By finding some
observables other than the MK polynomials that are less symmetric,
one may give a finer classification of $N$-qubit entanglement. It
will be of interest to find some criteria to discriminate such
degeneracy of the entanglement class.

This work was supported by the National Natural Science Foundation
of China, the Chinese Academy of Sciences and the National
Fundamental Research Program (under Grant No. 2001CB309303).


\begin{thebibliography}{99}
\bibitem{bell} J.S. Bell, Physics {\bf 1}, 165 (1964).
\bibitem{gisin1} N. Gisin, Phys. Lett. A {\bf 154}, 201 (1991).
\bibitem{mermin} N.D. Mermin, Phys. Rev. Lett. {\bf 65}, 1838 (1990).
\bibitem{gisin} N. Gisin and H. Bechmann-Pasquinucci, Phys. Lett. A
246, {\bf 1} (1998).
\bibitem{bruk} M. \.Zukowski and \v{C}.
Brukner, Phys. Rev. Lett. {\bf 88}, 210401 (2002).
\bibitem{pan} J.-W. Pan {\it et al.}, Phys. Rev. Lett. {\bf 86},
4435 (2001); C. A. Sackett {\it et al.}, Nature (London) {\bf
404}, 256 (2000).
\bibitem{kly}D. N. Klyshko, Phys. Lett. A {\bf 172}, 399 (1993);
A.V. Belinskii and D. N. Klyshko, Phys. Usp. {\bf 36}, 653
(1993).
\bibitem{svet1} G. Svetlichny, Phys. Rev. D {\bf 35}, 3066
(1987).
\bibitem{collins} D. Collins, N. Gisin, S. Popescu, D. Roberts, and V. Scarani
Phys. Rev. Lett. {\bf 88}, 170405 (2002).
\bibitem{svet2} M. Seevinck and G. Svetlichny, Phys. Rev. Lett.
{\bf 89}, 060401 (2002).
\bibitem{uffink} J. Uffink, Phys. Rev. Lett. {\bf 88}, 230406 (2002).
\bibitem{CHSH} J.F. Clauser, M.A. Horne, A. Shimony, and R.A. Holt,
Phys. Rev. Lett. {\bf 23},880 (1969).
\bibitem{sqrt2} B.S. Cirel'son, Lett. Math. Phys. {\bf 4}, 93
(1980).
\bibitem{ghz} D.M. Greenberger, M.A. Horne, A. Shimony, and A. Zeilinger,
Am. J. Phys. {\bf 58}, 1131 (1990).
\end{thebibliography}
\end{document}